\documentclass[9pt,twocolumn,twoside]{opticajnl}
\journal{opticajournal} 
\usepackage{siunitx}
\usepackage{bm}
\newcommand{\dif}{\mathrm{d}}
\setboolean{shortarticle}{true}
\newcommand{\myrev}[1]{#1}

\usepackage{lineno}

\title{Vortex inverted pin beams: Mitigation of scintillations in strong atmospheric turbulence}

\author[1,*]{Sotiris Droulias}
\author[2]{Michalis Loulakis}
\author[2,3]{Dimitris G. Papazoglou}
\author[2,3]{Stelios Tzortzakis}
\author[4]{Zhigang Chen}
\author[1,4,5,**]{Nikolaos K. Efremidis}

\affil[1]{Institute of Applied and Computational Mathematics, Foundation for Research and Technology - Hellas, 70013 Heraklion, Crete, Greece}
\affil[2]{Institute of Electronic Structure and Laser, Foundation for Research and Technology - Hellas, 70013, Heraklion, Crete, Greece}
\affil[3]{Department of Materials Science and Technology, University of Crete, 70013 Heraklion, Crete, Greece}
\affil[4]{MOE Key Laboratory of Weak-Light Nonlinear Photonics, TEDA Applied Physics Institute and School of Physics, Nankai University, Tianjin 300457, China}
\affil[5]{Department of Mathematics and Applied Mathematics, University of Crete, 70013 Heraklion, Crete, Greece}
\affil[*]{sdroulias@unipi.gr}
\affil[**]{nefrem@uoc.gr}

\begin{abstract}
  We recently introduced a new class of optical beams with a  Bessel-like transverse profile and increasing beam width during propagation, akin to an ``inverted pin''. Owing to their specially engineered distribution, these beams have shown remarkable performance in atmospheric turbulence. Specifically, inverted pin beams were found to have reduced scintillation index as compared to collimated or focused Gaussian beams as well as other types of pin beams especially in moderate to strong turbulence. In this work, we demonstrate that inverted pin beams carrying orbital angular momentum can further suppress intensity scintillations in moderate to strong irradiance fluctuation conditions. Our results can be useful in improving the performance and link availability of free-space optical communication systems. 
\end{abstract}

\setboolean{displaycopyright}{false} 

\begin{document}

\maketitle

The application of laser beam propagation over kilometric distances through the atmosphere for optical communications, can offer significant advantages over conventional radio frequency systems, such as higher bit rates, power concentration and security, and low latency~\cite{andre-2005}. However, the use of smaller wavelengths means that optical waves are less immune to weather conditions, such as absorption and scattering. Importantly, optical turbulence or fluctuations of the refractive index in the atmosphere result to intensity scintillations of the beam profile at the receiver. Such fluctuations might significantly impact the link quality and availability, resulting in link outages or degrade the link performance below acceptable levels.

The use of structured light beams has been actively investigated as a means to reduce irradiance fluctuations. In this respect, propagation invariant beams such as Bessel beams~\cite{eyyub-apb2007, eyyub-josaa2009, eyyub-ao2013} and Airy beams~\cite{efrem-optica2019, gu-ol2010, chu-ol2011, ji-oe2013, nelson-josaa2014} have been examined for their potential to reduce intensity scintillations. 

Optical waves with a topological charge, also known as optical vortices~\cite{coull-oc1989, bramb-pra1991}, have been extensively studied in optics~\cite{Shen-Light2019}. They are associated with an integer topological charge $n$, so that the phase continuously increases from $0$ to $2\pi n$ along a closed loop that surrounds the zero amplitude phase singularity. Optical vortices have been utilized in applications ranging, among others, from particle manipulation~\cite{pater-science2001, zhao-sr2015}, microscopy and imaging~\cite{tambu-prl2006}, as well as orbital angular momentum (OAM) multiplexing in optical communications~\cite{wang-pr2016, willn-apr2021}.
\myrev{In free-space optical communications, atmospheric turbulence can distort the OAM modes at the receiver introducing cross-talk between different modes~\cite{angui-ao2008}. Different mitigation methods at the receiver have been proposed to compensate for atmospheric turbulence~\cite{li-oc2018}.}
A variety of optical vortex beams have been studied for propagation in atmospheric turbulence, including Gaussian beams~\cite{gbur-josaa2008, gu-josaa2013} and their generalizations such as Laguerre-Gaussian beams and Hermite-Gaussian beams~\cite{zhu-oe2008, aksen-pr2015, fu-pr2016, lukin-pr2020}, vector vortex beams~\cite{gu-oe2020}, as well as the turbulent eigenmodes~\cite{krug-ap2023}. 

In 2019 a new class of pin-like structured light beams was introduced~\cite{zhang-aplp2019}, see also Refs.~\cite{gouts-ol2020, li-osac2020}. Importantly, it was experimentally demonstrated that pin beams (PBs) are far more resilient, as compared to standard Gaussian beams, after propagation of kilometric distance through atmospheric turbulence. Subsequently, such pin beams have been generalized in the form of vortex pin-like beams~\cite{bongi-pr2021, cao-josaa2022, xu-SPIE2023}. Recently, we proposed an inverted type of pin beam: A pin beam whose beam radius increases during propagation~\cite{droul-ol2023}, a broadening that is related to the beam structuring rather than natural diffraction. We showed that, in terms of the scintillation index (SI), inverted pin beams (IPBs) outperform the Gaussian (focused or collimated), regular pin, and Bessel beams. Such beams have been utilized to increase robustness and depth of focus in free-space optical communications~\cite{hu-ol2022, yang-jp2023}.

In this work, we investigate the dynamics of inverted pin beams with a OAM in atmospheric turbulence. We show that such vortex inverted pin (VIP) beams can have reduced SI as compared to regular pin beams when the irradiance fluctuations are moderate to strong. We numerically find the VIP beams that have optimum SI for different values of the topological charge, the refractive index structure parameter, and the receiver aperture. Our results show that the interplay between the receiver aperture, the size of the beam at the receiver, and the beam wander due to turbulence is particularly important in the optimum selection of the OAM. Furthermore, we plot the dynamics of the SI of the optimally selected VIP beams as a function of the propagation distance. We believe that our findings are promising for applications in free-space optical communications.

\begin{figure}[t!]
  \centering
  \includegraphics[width=\linewidth]{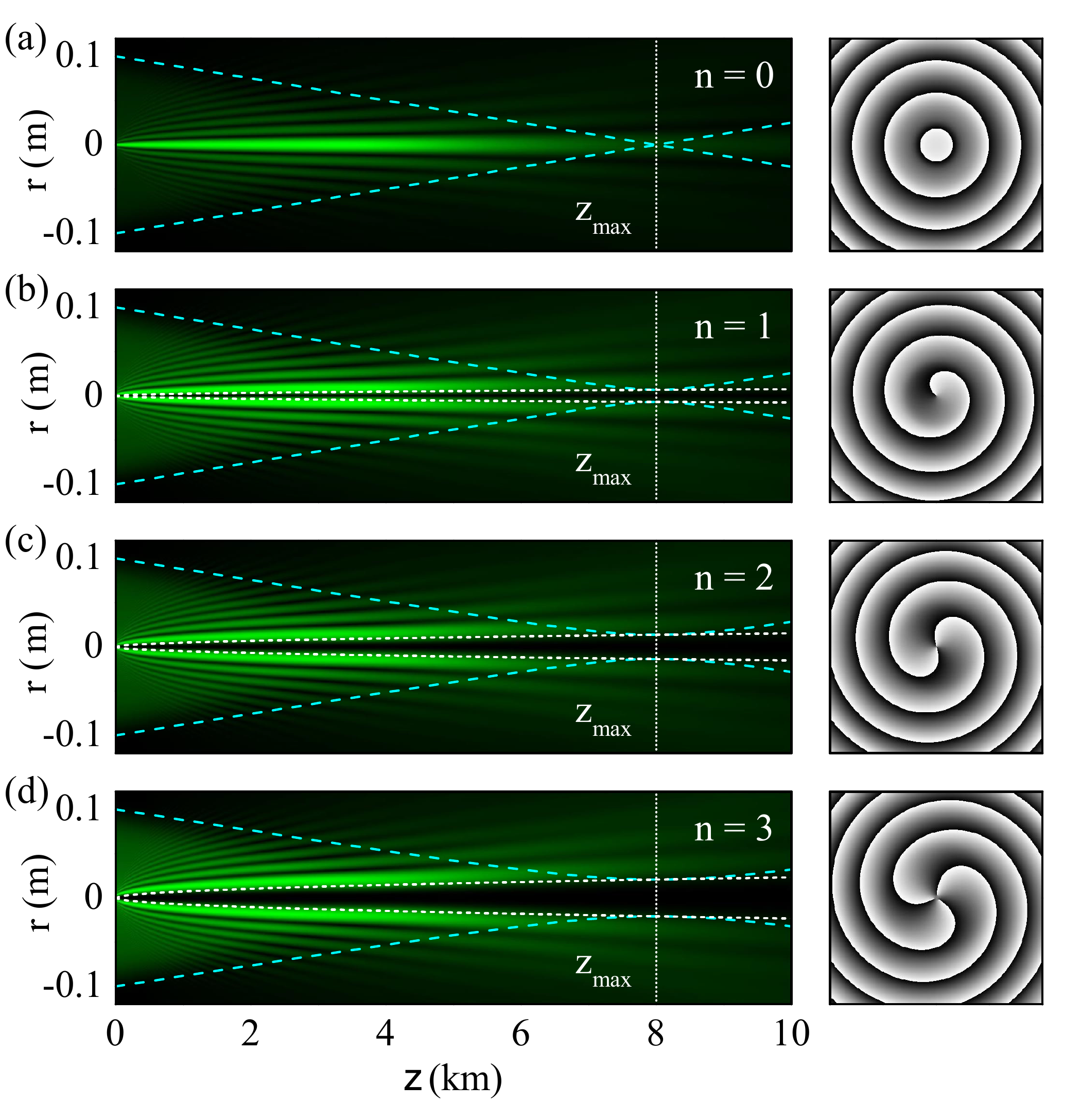}  
  \caption{Field amplitude propagation dynamics of VIP beams with $\gamma=0.4$ and topological charges 
    (a) $n = 0$,
    (b) $n = 1$,
    (c) $n = 2$, and
    (d) $n = 3$.
    In these examples $z_\mathrm{max}=8$ \unit{\kilo\meter} is the maximum propagation distance of the beams as defined in the text. The cyan dashed curves depict the radial trajectory of the rays [Eq.~(\ref{eq:ray})] that originate from the edge of the aperture $\rho=D_t/2$. The inner dotted curves are the focal coordinates as given by Eqs.~(\ref{eq:FocalCoordinates}). The panels on the right column show the phase of each beam after $L=6$ \unit{\kilo\meter} of propagation.
  }
  \label{fig:01}
\end{figure}

Let us consider the paraxial beam propagation as described by the Fresnel diffraction integral
\begin{equation}
  \psi(\bm r,z) = \frac{1}{i\lambda z}
  \iint_{\mathbb R^2}\psi_0(\bm\rho)
  \exp\left[
    \frac{ik}{2z}(\bm r-\bm\rho)^2
  \right]\dif\xi\dif\eta,
  \label{eq:Fresnel}
\end{equation}
where $z$ is the propagation distance, $\bm r=(x,y)$ are the transverse coordinates, which at the $z=0$ plane are defined as $\bm\rho=(\xi,\eta)$, $k=2\pi/\lambda$ is the wavenumber, and $\lambda$ the wavelength. In polar coordinates we can express $\bm r=(r,\theta)$ and $\bm\rho=(\rho,\varphi)$. The amplitude--phase decomposition of a vortex pin beam on the initial plane is given by~\cite{bongi-pr2021, gouts-ol2020}
\begin{equation}
  \psi_0(\rho,\varphi) = A(\rho)e^{i\Phi(\rho,\varphi)},
  \label{eq:psi0}
\end{equation}
where $\Phi(\rho,\varphi)=\phi(\rho)-n\varphi = -kC\rho^\gamma-n\varphi$, $n$ is the topological charge, $\gamma$ is the phase chirp, and $C$ controls the speed of phase variations. The equation that defines the radial distance of a ray from the origin $r(z)$ is~\cite{gouts-ol2020, bongi-pr2021}
\begin{equation}
  r^2=(\rho-C\gamma z\rho^{\gamma-1})^2+
  (nz/(k\rho))^2.
  \label{eq:ray}
\end{equation}
We define the propagation distance at which a ray is closest to the optical axis $r=r_f$ as the focal distance $z=z_f$. We can then derive the focal coordinates as a function of the radius of the ray on the initial plane $\rho$, 
\begin{equation}
  z_f = \frac{k^2\gamma C \rho^{\gamma+2}}{n^2+(k\gamma C \rho^{\gamma})^2},\quad
  r_f =
  \frac{|n|\rho}{(n^2+(k\gamma C \rho^{\gamma})^2)^{1/2}}.
  \label{eq:FocalCoordinates}
\end{equation}
By selecting in Eq.~(\ref{eq:FocalCoordinates}) $\rho$ to be equal to the transmitter radius $D_t/2$, then $z_f=z_\mathrm{max}$ is the maximum propagation distance of the VIP beam. We can utilize this relation to determine $C$ as a function of $D_t$ and $z_\mathrm{max}$, 
\begin{equation}
  C(D_t,z_\mathrm{max})
  =
  \frac{
    D_t^2 k+
    \sqrt{D_t^4 k^2-64n^2 z_\mathrm{max}^2}
  }{
    2^{3-\gamma}D_t^{\gamma}\gamma  kz_\mathrm{max}
  }.
  \label{eq:CDtzmax}
\end{equation}
If we further assume that $\rho\gg r_f(\rho)$ (or equivalently $\rho|\phi'(\rho)|\gg n$), then we can inverse Eq.~(\ref{eq:FocalCoordinates}) to directly determine $\rho(z_f)$. Thus, close to the axis, the following asymptotic formula is valid~\cite{bongi-pr2021}
\begin{multline}
  \psi(r, \theta, z)=
  |i|^{-n}\sqrt{2 \pi k/(2-\gamma)}
  (C \gamma z^{\frac{\gamma}{2}})^{\frac{1}{2-\gamma}}
  \\
  A[(C\gamma z)^{1/(2-\gamma)}]
  J_n[k r(C\gamma z^{\gamma-1})^{1/(2-\gamma)}]
  e^{i\Psi},
  \label{eq:psirz}
\end{multline}
where
$
\Psi = k r^2/(2z)+(C^2 \gamma^2 z^\gamma)^{1/(2-\gamma)} (k/2)(1-2/\gamma)-\pi/4(1+2|n|)-n\theta,
$
and we have replaced $z_f$ with $z$. 
An expression that does not rely on this approximation can also be found~\cite{gouts-ol2020}. In our simulations, the initial amplitude is selected so that the theoretically predicted maximum amplitude remains constant during propagation 
$A(\rho) = A_0\sqrt{(2-\gamma)/(2\pi\gamma k C)}i^{|n|}\rho^{-\gamma/2}.$
In agreement with numerical results, from Eq.~(\ref{eq:psirz}) we see that for $1<\gamma<2$ the vortex PB hollow core decreases with the propagation distance. For $\gamma=1$ the core remains constant during propagation resulting to a vortex Bessel beam. Finally, for $0<\gamma<1$ the hollow core increases with $z$ leading to the formation of an inverted vortex PB.

In this paper, we select a set of parameters that is similar to our previous work~\cite{droul-ol2023}: In particular, the optical wavelength is $\lambda=532$ \unit{\nano\meter}, the aperture diameter of the transmitter is $D_t=20$ \unit{\centi\meter}, and we focus on examining inverted pin beams with $\gamma=0.4$. In Fig.~\ref{fig:01}, we compare the dynamics of pin beams with different topological charges in the absence of turbulence. For $n=0$ the beam propagates as a typical inverted pin beam with its main lobe expanding along $z$. For $n\ge1$, as $n$ increases, the main lobe forms a vortex ring of increasing radius.
The panels on the right depict the phase of the beam after $L=6$ \unit{\kilo\meter} of propagation, with the $n^{\mathrm{th}}$ order phase singularity located at the center.

To assess the irradiance fluctuations, we separately optimize each beam in terms of the scintillation index
 $ \sigma_I^2=\langle I^2\rangle/\langle I\rangle^2-1,$
where $I=|\psi|^2$ is the beam irradiance and the brackets denote ensemble average. In our simulations we found that averaging over 1000 realizations results to accurate estimations for the SI. We also use the same screens for all types of the examined beams to ensure a fair comparison. We select a common receiver distance $L=6$ \unit{\kilo\meter} at which we want to minimize the SI of each beam. The maximum propagation distance of the pin beam, $z_\mathrm{max}$, is selected as the tuning parameter. The numerical methods and the respective parameters are described in detail in our previous work~\cite{droul-ol2023}. For the refractive index fluctuations, we use the modified von K\'arm\'an power spectral density model
$\Phi_n(\kappa) = 0.033 C_n^2
\exp{(-\kappa^2/\kappa_m^2)}
(\kappa^2 + \kappa_0^2)^{-11/6},$
where $\kappa_m=5.92 / l_0$, $\kappa_0=2\pi / L_0$ and $l_0$, $L_0$ are the inner and outer scales of the inertial subrange, respectively. Specifically, we select $l_0=5$ \unit{\milli\meter} and $L_0=10$ \unit{\meter}, which are typical values for propagation close to the surface of the earth~\cite{andre-2005}.

\begin{figure}[t!]
  \centering
  \includegraphics[width=\linewidth]{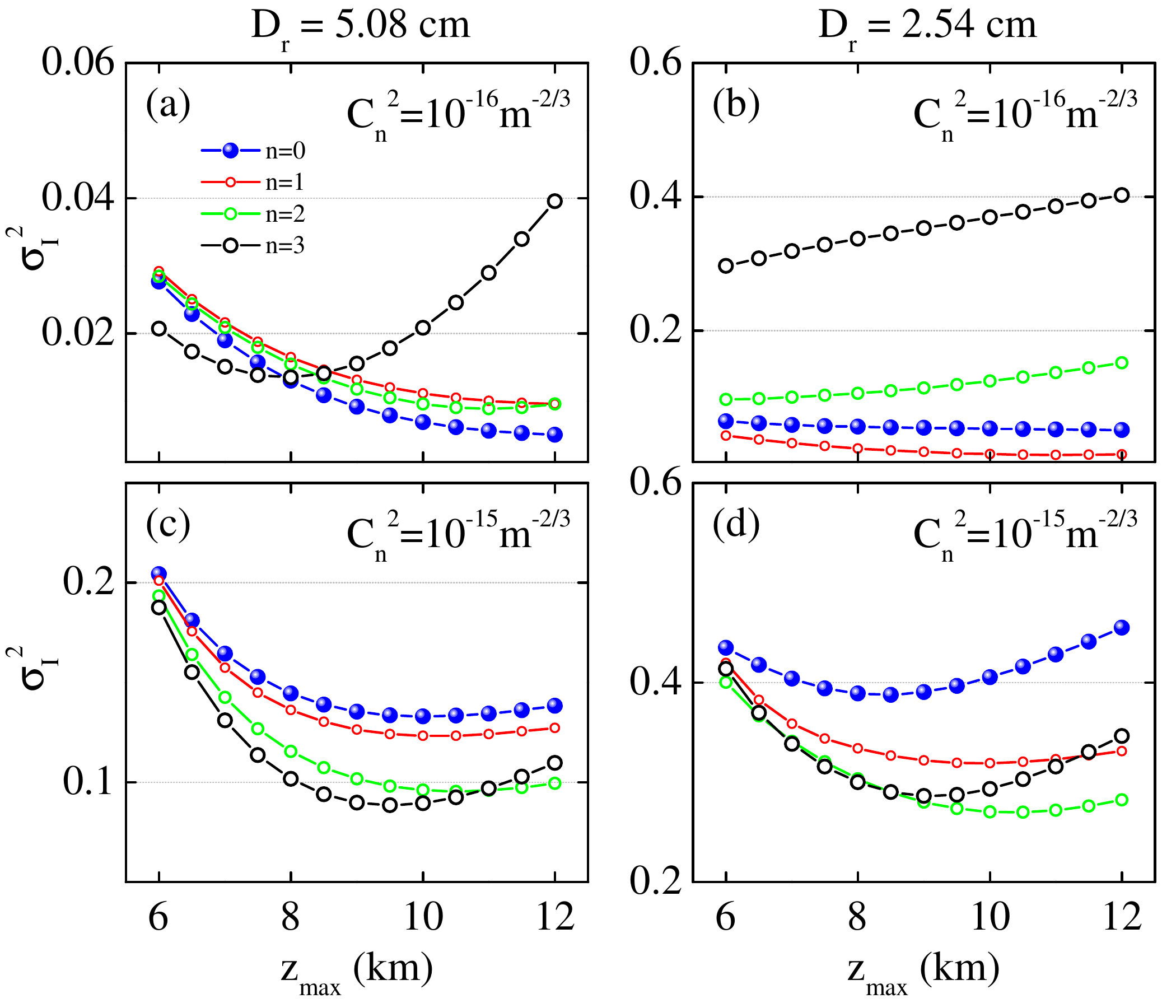}
  \caption{Optimal pin beam selection with $\gamma=0.4$ at receiver distance $L=6$ \unit{\kilo\meter}, based on the scintillation index for $C_n^2=10^{-16}$ \unit{\meter^{-2/3}} ($\sigma_R^2=0.185$) in first row and $C_n^2=10^{-15}$ \unit{\meter^{-2/3}} ($\sigma_R^2=1.85$) in the second row. The receiver aperture is
    $D_r=5.08$ \unit{\centi\meter}
    in the first column and
    $D_r=2.54$ \unit{\centi\meter}
    in the second column. Beam parameters are the same as in Fig.~\ref{fig:01}.}
  \label{fig:02}
\end{figure}

In Fig.~\ref{fig:02}, we individually optimize, in terms of the SI, the VIP beams with $\gamma=0.4$ and topological charge $n=0,1,2,3$ propagating over $L=6$ \unit{\kilo\meter} distance as a function of $z_\mathrm{max}$. We use four different parameter sets for weak ($C_n^2=10^{-16}$ \unit{\meter^{-2/3}} resulting to Rytov variance $\sigma_R^2=1.23 C_n^2 k^{7/6}L^{11/6}=0.185$) and strong ($C_n^2=10^{-15}$ \unit{\meter^{-2/3}}, $\sigma_R^2=1.85$) irradiance fluctuations and
$D_r=2.54$, $5.08$ \unit{\centi\meter}
receiver apertures.
Under optimal conditions [weak fluctuations and large receiver aperture shown in Fig.~\ref{fig:02}(a)], the smaller size of the $n=0$ IPB in conjunction with the limited amount of beam wander leads to smaller values of the SI when $z_\mathrm{max}=12$ \unit{\kilo\meter}. However, as the aperture diameter decreases or the
structure parameter $C_n^2$ increases, larger values of $n$ start to become favorable.
For example, for a smaller receiver aperture [Fig.~\ref{fig:02}(b)] the $n=1$ VIP beam  with $z_\mathrm{max}\approx11.5$ \unit{\kilo\meter} is optimal.
On the other hand, increasing the amount of turbulence to $C_n^2=10^{-15}$ \unit{\meter^{-2/3}} results to a minimization of the SI of the VIP beams with $n=3$ and $z_\mathrm{max}\approx10.5$ \unit{\kilo\meter}, as shown in Fig.~\ref{fig:02}(c).
Finally, as expected, the largest values of the SI are observed when both the receiver aperture is reduced and the intensity fluctuations are strong, as shown in Fig.~\ref{fig:02}(d). In this case, the optimal value of the topological charge is $n=2$ with $z_\mathrm{max}\approx10.5$ \unit{\kilo\meter}.
As $\sigma_I^2$ increases in the two columns of Fig.~\ref{fig:02}, we observe a shift of the optimum $z_\mathrm{max}$ to smaller values. Note that, although not shown in Fig.~\ref{fig:02}, in all cases examined here the $n=4$ VIP beams have larger optimal values of $\sigma_I^2$ as compared to the $n=3$. This is mainly attributed to the larger radius of the $n=4$ VIP beams.
We conclude that VIP beams have reduced SI in moderate to strong fluctuation conditions as compared to the $n=0$ IPB. The optimal value of the OAM depends on the environmental parameters and the setup. 

\begin{figure}[t!]
  \centering
  \includegraphics[width=\linewidth]{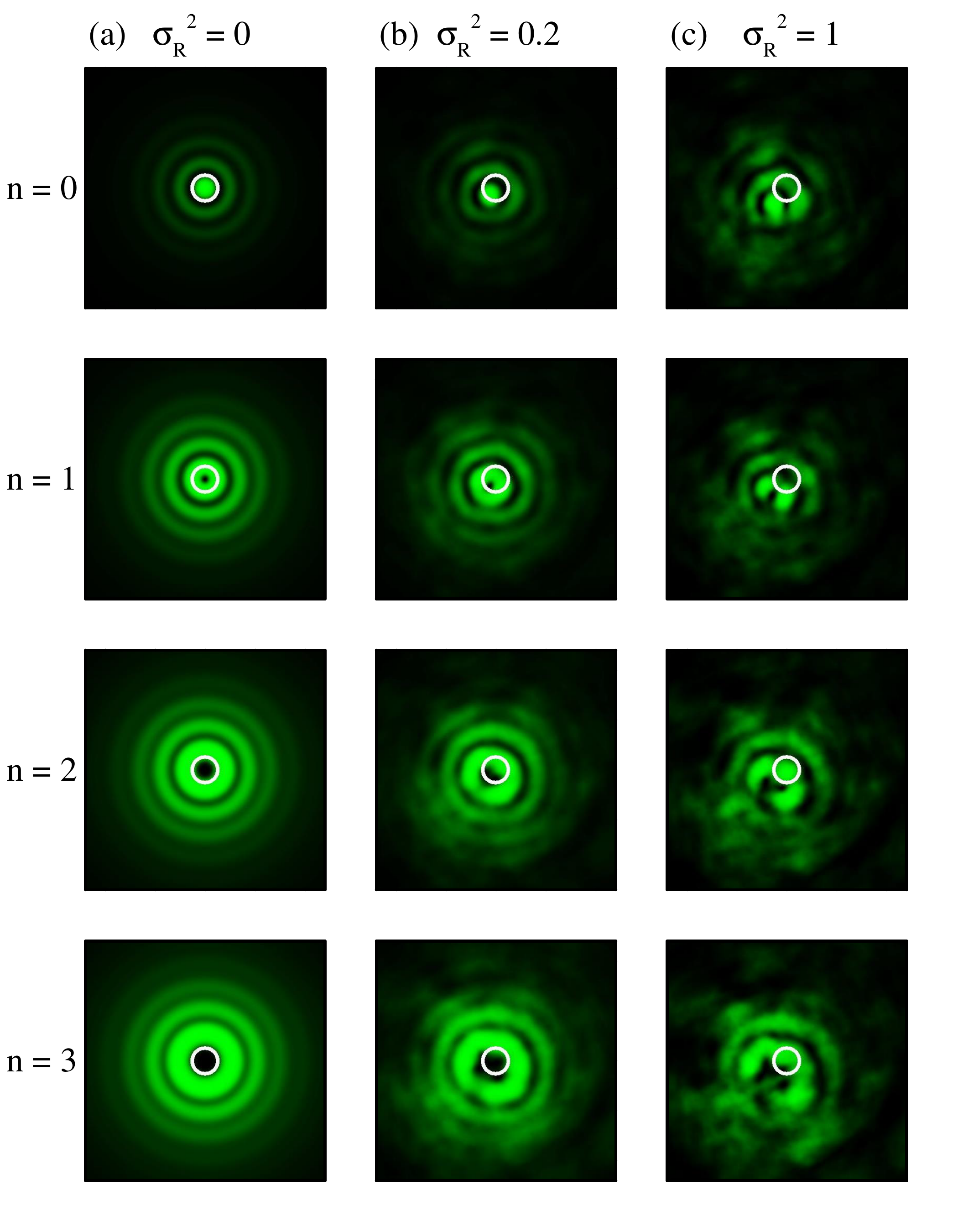}
  \caption{Intensity profile of VIP beam realizations with $\gamma=0.4$, after propagation of $L=6$ \unit{\kilo\meter} in a medium that leads to (a) no fluctuations, $\sigma_R^2=0$, (b) weak fluctuations, $\sigma_R^2=0.2$, and (c) moderate fluctuations, $\sigma_R^2=1$. The white circle marks the edge of the $1$ \unit{in} receiver aperture area. The beam parameters are the same as in Fig.~\ref{fig:01}.}
  \label{fig:03}
\end{figure}

To understand the underlying mechanisms in our results, in Fig.~\ref{fig:03} we plot the intensity profile of single VIP beam realizations after $6$ \unit{\kilo\meter} of propagation for (a) zero $\sigma_R^2=0$, (b) weak $\sigma_R^2=0.2$, and (c) moderate $\sigma_R^2=1$ irradiance fluctuations. In order to simulate identical turbulence conditions we use the same set of phase screens and vary only $C_n^2$.
We observe that the main factor that affects the performance of these beams is the interplay between three different length scales and, in particular, the beam radius, the receiver radius, and the amount of beam wander. For weak intensity fluctuations and, thus, small beam wander, narrower pin beams associated with smaller values of $n$, lead to output profiles that are at least partially located inside the receiver. Increasing $n$ results to a hollow core that can be as large or even larger than the receiver. Thus, even weak fluctuations can cause large variances in the amount of power captured among different realizations. Our observations are reversed in the case of moderate intensity fluctuations. In this case, due to increased beam wander, smaller values of $n$ can lead to a pin beam with a main lobe which is located outside the receiver. On the other hand, the increased ring size of VIP beams with larger $n$, contributes in receiving an appreciable amount of signal. 

\begin{figure}[t!]
  \centering
  \includegraphics[width=\linewidth]{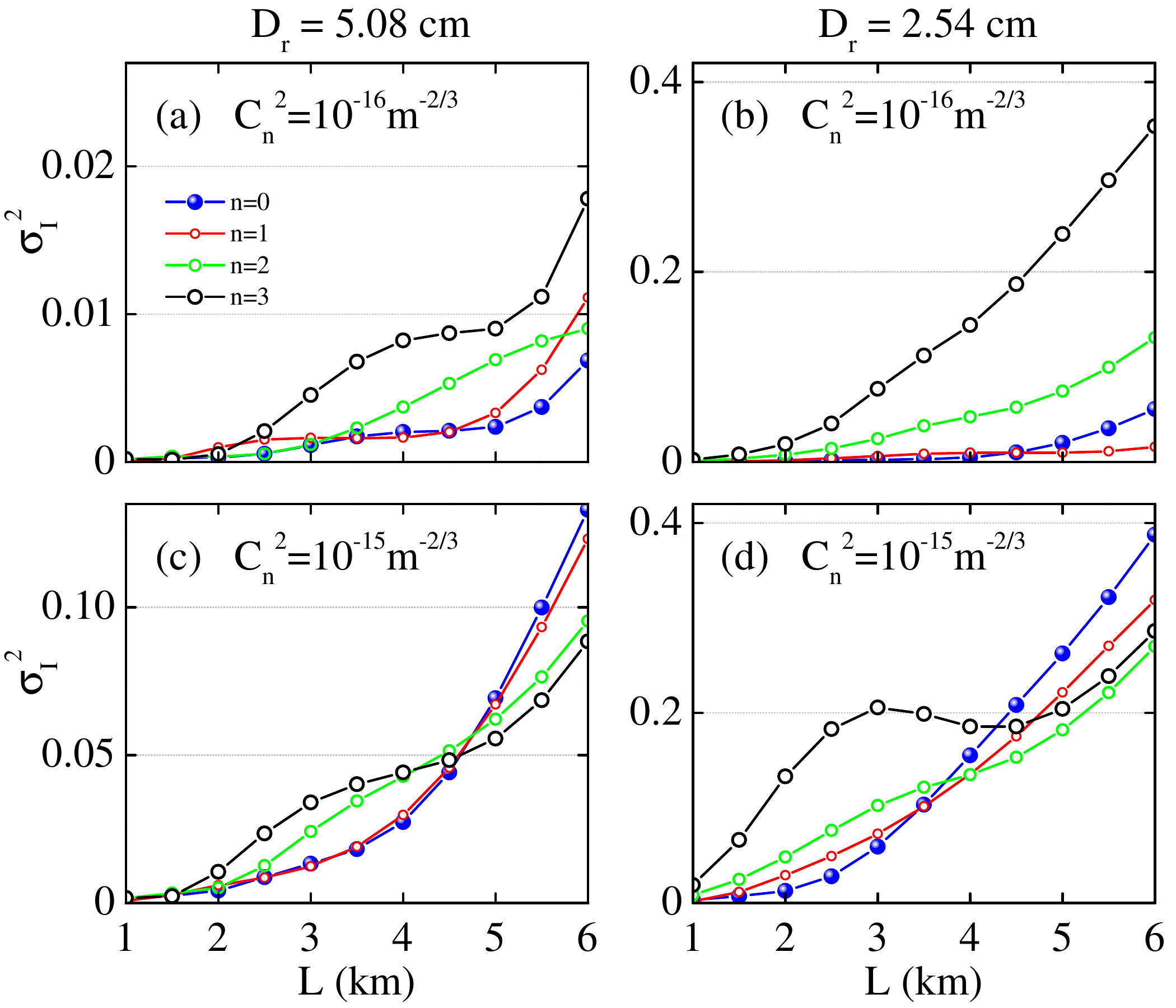}  
  \caption{SI of VIP beams with $\gamma=0.4$ as a function of the propagation distance $L$. In the first and second row $C_n^2=10^{-16}$  \unit{m^{-2/3}} ($\sigma_R^2=0.185$) and $C_n^2=10^{-15}$ \unit{m^{-2/3}} ($\sigma_R^2=1.85$) whereas in the first and second column the receiver aperture is
    $D_r=5.08$ \unit{\centi\meter} and $D_r=2.54$ \unit{\centi\meter},
    respectively. The beams in each column have been optimized for minimum SI when $C_n^2=10^{-15}$ \unit{\meter^{-2/3}}, according to the simulations of Fig.~\ref{fig:02}.}
  \label{fig:04}
\end{figure}

In Fig.~\ref{fig:04}, we select VIP beams with $n=\numrange{0}{3}$, which are optimized according to Fig.~\ref{fig:02} to have reduced SI under strong irradiance fluctuations $C_n^2=10^{-15}$ \unit{\meter^{-2/3}} (with $L=6$ \unit{\kilo\meter} and
$D_r=2.54$ \unit{\centi\meter} or $5.08$ \unit{\centi\meter}).
The performance of these beams is then tested under different turbulence strengths $C_n^2=10^{-16}$ \unit{\meter^{-2/2}} and $C_n^2=10^{-15}$ \unit{\meter^{-2/3}} for a range of propagation distances $L=\numrange{1}{6}$ \unit{\kilo\meter}.
When
$D_r=5.08$ \unit{\centi\meter}
and $C_n^2=10^{-15}$ \unit{\meter^{-2/3}}, the pin beam with $n=3$ has the minimum SI according to Fig.~\ref{fig:02}, closely followed by the $n=2$. However, for smaller propagation distances the $n=0,1$ become optimum [Fig.~\ref{fig:04}(c)].
Reducing turbulence strength to $C_n^2=10^{-16}$, as shown in Fig.~\ref{fig:04}(a), the fundamental $n=0$ pin beam has better performance for most of the propagation distances examined.
The same trend
is also observed in Figs.~\ref{fig:04}(b), (d). Specifically, in Fig.~\ref{fig:04}(b) $n=1\rightarrow n=0$, whereas in Fig.~\ref{fig:04}(d) $n=2\rightarrow n=1\rightarrow n=0$, as $L$ decreases. 
The overall conclusion is that for smaller propagation distances, and thus weaker irradiance fluctuations, the smaller values of the topological charge have a better performance. However, as the intensity fluctuations become stronger, higher topological charges display significantly better performance.
\myrev{In Supplementary material, we examine the SI of PBs that are optimized for specific values of $C_n^2$ and $D_r$, as either one of them varies. These results are in agreement with the examples shown in the main text, showing better performance of VIP beams with $n=2,3$ in the moderate to strong intensity fluctuation regime.}

In conclusion, we have investigated the scintillation index of vortex inverted pin beams. We find that such beams can have better performance than regular ($n=0$) pin beams in the case of moderate to strong intensity fluctuations. These results build up and expand our recently published work, where we showed that regular pin beams have a significant advantage in comparison to collimated and focused Gaussian beams~\cite{droul-ol2023}. The simplicity in the generation in conjunction with the improved performance of VIP beams can make such optical waves useful in free-space optical communication systems~\cite{willn-apr2021}. 

\begin{backmatter}
  
  \bmsection{Funding} This work was supported by Huawei Technologies Sweden AB.
  
  \bmsection{Acknowledgments} The authors would like to thank Ulrik Imberg from Huawei for his assistance in this project and Domenico Bongiovanni for discussions.

  \bmsection{Disclosures} The authors declare no conflicts of interest.

  \bmsection{Data Availability Statement} Data underlying the results presented in this paper are not publicly available at this time but may be obtained from the authors upon reasonable request.

  \bmsection{Supplemental document} See Supplement 1 for supporting content.

\end{backmatter}

\newcommand{\noopsort[1]}{} \newcommand{\singleletter}[1]{#1}

\newpage

\renewcommand\refname{FULL REFERENCES}


\begin{thebibliography}{10}
\newcommand{\enquote}[1]{``#1''}

\bibitem{andre-2005}
L.~C. Andrews and R.~L. Phillips, \emph{Laser Beam Propagation Through Random
  Media} (SPIE press, Bellingham, Washington USA, 2005), 2nd ed.

\bibitem{eyyub-apb2007}
H.~T. Eyyubo{\u{g}}lu, \enquote{Propagation of higher order bessel--gaussian
  beams in turbulence,} {\protect\JournalTitle{Applied Physics B}} \textbf{88},
  259--265 (2007).

\bibitem{eyyub-josaa2009}
H.~T. Eyyubo\u{g}lu, Y.~Baykal, E.~Sermutlu, O.~Korotkova, and Y.~Cai,
  \enquote{Scintillation index of modified bessel-gaussian beams propagating in
  turbulent media,} {\protect\JournalTitle{J. Opt. Soc. Am. A}} \textbf{26},
  387--394 (2009).

\bibitem{eyyub-ao2013}
H.~T. Eyyubo\u{g}lu, D.~Voelz, and X.~Xiao, \enquote{Scintillation analysis of
  truncated bessel beams via numerical turbulence propagation simulation,}
  {\protect\JournalTitle{Appl. Opt.}} \textbf{52}, 8032--8039 (2013).

\bibitem{efrem-optica2019}
N.~K. Efremidis, Z.~Chen, M.~Segev, and D.~N. Christodoulides, \enquote{Airy
  beams and accelerating waves: an overview of recent advances,}
  {\protect\JournalTitle{Optica}} \textbf{6}, 686--701 (2019).

\bibitem{gu-ol2010}
Y.~Gu and G.~Gbur, \enquote{Scintillation of {A}iry beam arrays in atmospheric
  turbulence,} {\protect\JournalTitle{Opt. Lett.}} \textbf{35}, 3456--3458
  (2010).

\bibitem{chu-ol2011}
X.~Chu, \enquote{Evolution of an {A}iry beam in turbulence,}
  {\protect\JournalTitle{Opt. Lett.}} \textbf{36}, 2701--2703 (2011).

\bibitem{ji-oe2013}
X.~Ji, H.~T. Eyyubo\u{g}lu, G.~Ji, and X.~Jia, \enquote{Propagation of an airy
  beam through the atmosphere,} {\protect\JournalTitle{Opt. Express}}
  \textbf{21}, 2154--2164 (2013).

\bibitem{nelson-josaa2014}
W.~Nelson, J.~P. Palastro, C.~C. Davis, and P.~Sprangle, \enquote{Propagation
  of bessel and airy beams through atmospheric turbulence,}
  {\protect\JournalTitle{J. Opt. Soc. Am. A}} \textbf{31}, 603--609 (2014).

\bibitem{coull-oc1989}
P.~Coullet, L.~Gil, and F.~Rocca, \enquote{Optical vortices,}
  {\protect\JournalTitle{Optics Communications}} \textbf{73}, 403--408 (1989).

\bibitem{bramb-pra1991}
M.~Brambilla, F.~Battipede, L.~A. Lugiato, V.~Penna, F.~Prati, C.~Tamm, and
  C.~O. Weiss, \enquote{Transverse laser patterns. i. phase singularity
  crystals,} {\protect\JournalTitle{Phys. Rev. A}} \textbf{43}, 5090--5113
  (1991).

\bibitem{Shen-Light2019}
Y.~Shen, X.~Wang, Z.~Xie, C.~Min, X.~Fu, Q.~Liu, M.~Gong, and X.~Yuan,
  \enquote{Optical vortices 30 years on: {OAM} manipulation from topological
  charge to multiple singularities,} {\protect\JournalTitle{Light: Science \&
  Applications}} \textbf{8}, 90 (2019).

\bibitem{pater-science2001}
L.~Paterson, M.~P. MacDonald, J.~Arlt, W.~Sibbett, P.~E. Bryant, and
  K.~Dholakia, \enquote{Controlled {Rotation} of {Optically} {Trapped}
  {Microscopic} {Particles},} {\protect\JournalTitle{Science}} \textbf{292},
  912--914 (2001).

\bibitem{zhao-sr2015}
J.~Zhao, I.~D. Chremmos, D.~Song, D.~N. Christodoulides, N.~K. Efremidis, and
  Z.~Chen, \enquote{Curved singular beams for three-dimensional particle
  manipulation,} {\protect\JournalTitle{Sci. Rep.}} \textbf{5}, 12086 (2015).

\bibitem{tambu-prl2006}
F.~Tamburini, G.~Anzolin, G.~Umbriaco, A.~Bianchini, and C.~Barbieri,
  \enquote{Overcoming the rayleigh criterion limit with optical vortices,}
  {\protect\JournalTitle{Phys. Rev. Lett.}} \textbf{97}, 163903 (2006).

\bibitem{wang-pr2016}
J.~Wang, \enquote{Advances in communications using optical vortices,}
  {\protect\JournalTitle{Photon. Res.}} \textbf{4}, B14--B28 (2016).

\bibitem{willn-apr2021}
A.~E. Willner, K.~Pang, H.~Song, K.~Zou, and H.~Zhou, \enquote{{Orbital angular
  momentum of light for communications},} {\protect\JournalTitle{Applied
  Physics Reviews}} \textbf{8}, 041312 (2021).

\bibitem{angui-ao2008}
J.~A. Anguita, M.~A. Neifeld, and B.~V. Vasic, \enquote{Turbulence-induced
  channel crosstalk in an orbital angular momentum-multiplexed free-space
  optical link,} {\protect\JournalTitle{Appl. Opt.}} \textbf{47}, 2414--2429
  (2008).

\bibitem{li-oc2018}
S.~Li, S.~Chen, C.~Gao, A.~E. Willner, and J.~Wang, \enquote{Atmospheric
  turbulence compensation in orbital angular momentum communications: Advances
  and perspectives,} {\protect\JournalTitle{Optics Communications}}
  \textbf{408}, 68--81 (2018). Optical Communications Exploiting the Space
  Domain.

\bibitem{gbur-josaa2008}
G.~Gbur and R.~K. Tyson, \enquote{Vortex beam propagation through atmospheric
  turbulence and topological charge conservation,} {\protect\JournalTitle{J.
  Opt. Soc. Am. A}} \textbf{25}, 225--230 (2008).

\bibitem{gu-josaa2013}
Y.~Gu, \enquote{Statistics of optical vortex wander on propagation through
  atmospheric turbulence,} {\protect\JournalTitle{J. Opt. Soc. Am. A}}
  \textbf{30}, 708--716 (2013).

\bibitem{zhu-oe2008}
K.~Zhu, G.~Zhou, X.~Li, X.~Zheng, and H.~Tang, \enquote{Propagation of
  bessel-gaussian beams with optical vortices in turbulent atmosphere,}
  {\protect\JournalTitle{Opt. Express}} \textbf{16}, 21315--21320 (2008).

\bibitem{aksen-pr2015}
V.~P. Aksenov and V.~V. Kolosov, \enquote{Scintillations of optical vortex in
  randomly inhomogeneous medium,} {\protect\JournalTitle{Photon. Res.}}
  \textbf{3}, 44--47 (2015).

\bibitem{fu-pr2016}
S.~Fu and C.~Gao, \enquote{Influences of atmospheric turbulence effects on the
  orbital angular momentum spectra of vortex beams,}
  {\protect\JournalTitle{Photon. Res.}} \textbf{4}, B1--B4 (2016).

\bibitem{lukin-pr2020}
I.~P. Lukin, \enquote{Coherence of vortex bessel-like beams in a turbulent
  atmosphere,} {\protect\JournalTitle{Appl. Opt.}} \textbf{59}, 3833--3841
  (2020).

\bibitem{gu-oe2020}
X.~Gu, L.~Chen, and M.~Krenn, \enquote{Phenomenology of complex structured
  light in turbulent air,} {\protect\JournalTitle{Opt. Express}} \textbf{28},
  11033--11050 (2020).

\bibitem{krug-ap2023}
A.~Klug, C.~Peters, and A.~Forbes, \enquote{{Robust structured light in
  atmospheric turbulence},} {\protect\JournalTitle{Advanced Photonics}}
  \textbf{5}, 016006 (2023).

\bibitem{zhang-aplp2019}
Z.~Zhang, X.~Liang, M.~Goutsoulas, D.~Li, X.~Yang, S.~Yin, J.~Xu, D.~N.
  Christodoulides, N.~K. Efremidis, and Z.~Chen, \enquote{Robust propagation of
  pin-like optical beam through atmospheric turbulence,}
  {\protect\JournalTitle{APL Photonics}} \textbf{4}, 076103 (2019).

\bibitem{gouts-ol2020}
M.~Goutsoulas, D.~Bongiovanni, D.~Li, Z.~Chen, and N.~K. Efremidis,
  \enquote{Tunable self-similar bessel-like beams of arbitrary order,}
  {\protect\JournalTitle{Opt. Lett.}} \textbf{45}, 1830--1833 (2020).

\bibitem{li-osac2020}
D.~Li, D.~Bongiovanni, M.~Goutsoulas, S.~Xia, Z.~Zhang, Y.~Hu, D.~Song,
  R.~Morandotti, N.~K. Efremidis, and Z.~Chen, \enquote{Direct comparison of
  anti-diffracting optical pin beams and abruptly autofocusing beams,}
  {\protect\JournalTitle{OSA Continuum}} \textbf{3}, 1525--1535 (2020).

\bibitem{bongi-pr2021}
D.~Bongiovanni, D.~Li, M.~Goutsoulas, H.~Wu, Y.~Hu, D.~Song, R.~Morandotti,
  N.~K. Efremidis, and Z.~Chen, \enquote{Free-space realization of tunable
  pin-like optical vortex beams,} {\protect\JournalTitle{Photon. Res.}}
  \textbf{9}, 1204--1212 (2021).

\bibitem{cao-josaa2022}
J.~Cao, L.~Han, H.~Liang, G.~Wu, and X.~Pang, \enquote{Orbital angular momentum
  spectrum of pin-like optical vortex beams in turbulent atmosphere,}
  {\protect\JournalTitle{J. Opt. Soc. Am. A}} \textbf{39}, 1414--1419 (2022).

\bibitem{xu-SPIE2023}
Y.~Xu, B.~Lan, C.~Liu, M.~Chen, H.~Sun, Y.~Zhang, and H.~Xian,
  \enquote{Self-focusing pin-like optical vortex beams resist atmospheric
  turbulence propagation for the space optical communication,} in \emph{3rd
  International Conference on Laser, Optics, and Optoelectronic Technology
  (LOPET 2023),}  vol. 12757 (SPIE, 2023), pp. 601--607.

\bibitem{droul-ol2023}
S.~Droulias, M.~Loulakis, D.~G. Papazoglou, S.~Tzortzakis, Z.~Chen, and N.~K.
  Efremidis, \enquote{Inverted pin beams for robust long-range propagation
  through atmospheric turbulence,} {\protect\JournalTitle{Opt. Lett.}}
  \textbf{48}, 5467--5470 (2023).

\bibitem{hu-ol2022}
N.~Hu, H.~Zhou, R.~Zhang, H.~Song, K.~Pang, K.~Zou, H.~Song, X.~Su, C.~Liu,
  B.~Lynn, M.~Tur, and A.~E. Willner, \enquote{Experimental demonstration of a
  ``pin-like'' low-divergence beam in a 1-gbit/s ook fso link using a
  limited-size receiver aperture at various propagation distances,}
  {\protect\JournalTitle{Opt. Lett.}} \textbf{47}, 4215--4218 (2022).

\bibitem{yang-jp2023}
Y.~Yang, X.~Kang, and L.~Cao, \enquote{Robust propagation of a steady optical
  beam through turbulence with extended depth of focus based on spatial light
  modulator,} {\protect\JournalTitle{Journal of Physics: Photonics}}
  \textbf{5}, 035002 (2023).

\end{thebibliography}
\end{document}